\documentclass[twocolumn,floatfix, showpacs]{revtex4}

\usepackage[final]{epsfig}
\usepackage{amsmath}
\usepackage{parskip}
 
\begin{document}
 
\title{A comparison study of medium-modified QCD shower evolution scenarios}
 
\author{Thorsten Renk}
\email{trenk@phys.jyu.fi}
%\author{Kari J.~Eskola}
%\email{kari.eskola@phys.jyu.fi}
\affiliation{Department of Physics, P.O. Box 35 FI-40014 University of Jyv\"askyl\"a, Finland}
\affiliation{Helsinki Institute of Physics, P.O. Box 64 FI-00014, University of Helsinki, Finland}
 
%\preprint{HIP-2006-46/TH}

\begin{abstract}
The computation of hard processes in hadronic collisions is a major success of perturbative Quantum Chromodynamics (pQCD). In such processes, pQCD not only predicts the hard reaction itself, but also the subsequent evolution in terms of parton branching and radiation, leading to a parton shower and ultimately to an observable jet of hadrons. If the hard process occurs in a heavy-ion collision, a large part of this evolution takes place in the soft medium created along with the hard reaction. An observation of jets in heavy-ion collision thus allows a study of medium-modified QCD shower evolution. In vacuum,  Monte-Carlo (MC) simulations are well established tools to describe such showers. For jet studies in heavy-ion collisions, MC models for in-medium showers are currently being developed. However, the shower-medium interaction depends on the nature of the microscopic degrees of freedom of the medium created in a heavy-ion collision which is the very object one would like to investigate. This paper presents a study in comparison between three different possible implementations for the shower-medium interaction, two of them based on medium-induced pQCD radiation, one of them a medium-induced drag force, and shows for which observables differences between the three scenarios become visible. We find that while single hadron observables such as $R_{AA}$ are incapable of differentiating between the scenarios, jet observables such as the longiudinal momentum spectrum of hadrons in the jet show the potential to do so.
\end{abstract}
 
\maketitle

\section{Introduction}

Jet quenching, i.e.\ the energy loss of hard partons created in the first moments of a heavy ion collision due to interactions with the surrounding soft medium  has long been regarded a promising tool to study properties of the soft medium \cite{Jet1,Jet2,Jet3,Jet4,Jet5,Jet6}. The basic idea is to study the changes induced by the medium to a hard process which is well-known from p-p collisions. A number of observables is available for this purpose, among them suppression in single inclusive hard hadron spectra $R_{AA}$ \cite{PHENIX_R_AA}, the suppression of back-to-back correlations \cite{Dijets1,Dijets2}, single hadron suppression as a function of the emission angle with the reaction plane \cite{PHENIX-RP} and most recently also preliminary measurements of jets have become available \cite{STARJET}.

Single hadron observables and back-to-back correlations are well described in detailed model calculations using the concept of energy loss \cite{HydroJet1,Dihadron1,Dihadron2}, i.e. under the assumption that the process can be described by a medium-induced shift of the leading parton energy by an amount $\Delta E$, followed by a fragmentation process using vacuum fragmentation of a parton with the reduced energy. However, there are also calculations for these observables in which the evolution of the whole in-medium parton shower is followed in an analytic way \cite{HydroJet2,HydroJet3,Dihadron3}. Recently, also Monte Carlo (MC) codes for in-medium shower evolution have become available \cite{JEWEL,YAS,YAS2,Carlos} which are based on MC shower simulations developed for hadronic collisions, such as PYTHIA \cite{PYTHIA} or HERWIG \cite{HERWIG}. In medium-modified shower computations, energy is not simply lost but redistributed in a characteristic way. 

All current in-medium shower MC codes model the interaction between partons and the medium in a different way. JEWEL (Jet Evolution With Energy Loss) \cite{JEWEL} assumes either elastic collisions with thermal quasiparticles or, to implement radiative energy loss, an enhancement of the singular part of the parton branching kernels. YaJEM (Yet another Jet Energy-loss Model) \cite{YAS,YAS2} makes the assumption that the virtuality of partons traversing the medium grows according to the medium transport coefficient $\hat{q}$ which measures the virtuality gain per unit pathlength, and this medium-induced virtuality leads to increased radiation. Finally, Q-PYTHIA, the code presented in \cite{Carlos} is a direct extension of the leading parton energy loss computations done in \cite{Jet4,QuenchingWeights} and uses the differential radiation probabilities originally computed from a single hard parton now for each parton propagating in the shower simulation. At this stage, it is hardly surprising that different models employ different implementations of the parton-medium interaction, as the nature of this interaction crucially depends on the microscopic properties of the medium, i.e. the very thing one wishes to determine from the experiments. 

A suitable strategy to determine these properties is thus to study the effects of various different implementations of the parton-medium interaction for different observables. In this paper, we begin such a program by investigating the effects on a number of different observables resulting from three different scenarios: Medium-induced radiation by an increase of parton virtuality dependent on the medium $\hat{q}$ as used in \cite{YAS,YAS2}, an enhancement of the singular parts of the branching kernel leading to additional radiation as used in \cite{JEWEL,HBP} and a drag force. Momentum-dependent drag forces appear in computations modelling QCD-like $N=4$ super Yang-Mills theories via the AdS/CFT conjecture \cite{AdS}, in the present paper we use a simplified ansatz in which a parton in a constant medium undergoes a momentum independent energy loss per unit pathlength. Such a drag term has not been tested in an in-medium shower evolution MC code previously. 

The paper is organized as follows: First, we briefly review the computation of medium-modified hadron jet as done in \cite{YAS,YAS2}. In addition, we describe the three different implementations of the parton-medium interaction and its relation to the spacetime structure of the shower in detail. In a first comparison, we make the connection to previous leading parton energy loss calculations by considering a constant medium with fixed length. In this medium, we study the energy loss of the leading parton and present the result in terms of energy loss probability distributions and mean energy loss as a function of the parameters characterizing the medium. In a second comparison, we turn to a  medium model which is closer to the experimental situation in so far as it expands and hence dilutes as a function of time. We compute various jet observables in this medium, such as the longitudinal momentum distribuion inside the jet or the angular broadening. Finally, in a last comparison we compute (as done in \cite{YAS}) the suppression of the inclusive single hard hadron spectrum in terms of the nuclear suppression factor $R_{AA}$ and compare all scenarios  with experimental results \cite{PHENIX_R_AA}. From this comparison, we tentatively deduce the relevant medium parameters.  We conclude with a discussion of the implications of the results.

\section{Medium-modified shower evolution}
\label{S-MMFF}

In this section, we describe how the medium-modified fragmentation function (MMFF) is obtained from a computation of an in-medium shower followed by hadronization. Key ingredient for this computation is a pQCD MC shower algorithm. In this work, we employ a modification of the PYTHIA shower algorithm PYSHOW \cite{PYSHOW}. In the absence of any medium effects, our algorithm therefore corresponds directly to the PYTHIA shower. Furthermore, the subsequent hadronization of the shower is assumed to take place outside of the medium, even if the shower itself was medium-modified. It is computed using the Lund string fragmentation scheme \cite{Lund} which is also part of PYTHIA.

\subsection{Shower evolution in vacuum}

We model the evolution from some initial, highly virtual parton to a final state parton shower as a series of branching processes $a \rightarrow b+c$ where $a$ is called the parent parton and $b$ and $c$ are referred to as daughters. 
 In QCD, the allowed branching processes are $q \rightarrow qg$, $g \rightarrow gg$ and $g \rightarrow q \overline{q}$.  The kinematics of a branching is described in terms of the virtuality scale $Q^2$ and of the energy fraction $z$, where the energy of daughter $b$ is given by $E_b = z E_a$ and of the daughter $c$ by $E_c = (1-z) E_a$. It is convenient to introduce $t = \ln Q^2/\Lambda_{QCD}$ where $\Lambda_{QCD}$ is the scale parameter of QCD. $t$ takes a role similar to a time in the evolution equations, as it describes the evolution from some high initial virtuality $Q_0$ ($t_0$) to a lower virtuality $Q_m$ ($t_m$) at which the next branching occurs. In terms of the two variables, the differential probability $dP_a$ for a parton $a$ to branch is \cite{DGLAP1,DGLAP2}

\begin{equation}
dP_a = \sum_{b,c} \frac{\alpha_s}{2\pi} P_{a\rightarrow bc}(z) dt dz
\end{equation}

where $\alpha_s$ is the strong coupling and the splitting kernels $P_{a\rightarrow bc}(z)$ read

\begin{eqnarray}
&&P_{q\rightarrow qg}(z) = 4/3 \frac{1+z^2}{1-z} \label{E-qqg}\\
&&P_{g\rightarrow gg}(z) = 3 \frac{(1-z(1-z))^2}{z(1-z)}\label{E-ggg}\\
&&P_{g\rightarrow q\overline{q}}(z) = N_F/2 (z^2 + (1-z)^2)\label{E-gqq}
\end{eqnarray}

where we do not consider electromagnetic branchings. $N_F$ counts the number of active quark flavours for given virtuality. 

At a given value of the scale $t$, the differential probability for a branching to occur is given by the integral over all allowed values of $z$ in the branching kernel as

\begin{equation}
I_{a\rightarrow bc}(t) = \int_{z_-(t)}^{z_+(t)} dz \frac{\alpha_s}{2\pi} P_{a\rightarrow bc}(z).
\end{equation}

The kinematically allowed range of $z$ is given by

\begin{widetext}
\begin{equation}
\label{E-KB}
z_\pm = \frac{1}{2} \left( 1+ \frac{M_b^2 - M_c^2}{M_a^2}\pm \frac{|{\bf p}_a|}{E_a}\frac{\sqrt{(M_a^2-M_b^2-M_c^2)^2 -4M_b^2M_c^2}}{M_a^2} \right)
\end{equation}
\end{widetext}

where $M_i^2 = Q_i^2 + m_i^2$ with $m_i$ the bare quark mass or zero in the case of a gluon. 
Given the initial parent virtuality $Q_a^2$ or equivalently $t_a$, the virtuality at which the next branching occurs can be determined with the help of the Sudakov form factor $S_a(t)$, i.e. the probability that no branching occurs between $t_0$ and $t_m$, where

\begin{equation}
S_a(t) = \exp\left[ - \int_{t_0}^{t_m} dt' \sum_{b,c} I_{a \rightarrow bc}(t') \right].
\end{equation}

Thus, the probability density that a branching of $a$ occurs at $t_m$ is given by

\begin{equation}
\label{E-Qsq}
\frac{dP_a}{dt} = \left[\sum_{b,c}I_{a\rightarrow bc}(t)  \right] S_a(t).
\end{equation}

These equations are solved for each branching by the PYSHOW algorithm \cite{PYSHOW} iteratively to generate a shower. For each branching first Eq.~(\ref{E-Qsq}) is solved to determine the scale of the next branching, then Eqs.~(\ref{E-qqg})-(\ref{E-gqq}) are evaluated to determine the type of branching and the value of $z$, if the value of $z$ is outside the kinematic bound given by Eq.~(\ref{E-KB}) then the event is rejected. Given $t_0, t_m$ and $z$, energy-momentum conservation determines the rest of the kinematics except for a radial angle by which the plane spanned by the vectors of the daughter parents can be rotated.

In order to account in a schematic way for higher order interference terms, angular ordering is enforced onto the shower, i.e. opening angles spanned between daughter pairs $b,c$ from a parent $a$ are enforced to decrease according to the condition

\begin{equation}
\label{E-Angular}
\frac{z_b (1-z)b)}{M_b^2} > \frac{1-z_a}{z_a M_a^2}
\end{equation}

After a branching process has been computed, the same algorithm is applied to the two daughter partons treating them as new mothers. The branching is continued down to a scale $Q_{min}$ which is set to 1 GeV in the MC simulation, after which the partons are set on-shell, adjusting transverse momentum to ensure energy-momentum conservation.

After all possible branchings have been performed, i.e. after for all partons the condition $Q\le Q_{min}$ has been reached, the resulting parton shower is connected with a string following the Lund scheme \cite{Lund} which is subsequently allowed to decay into hadrons. These hadrons form the observable jet, and analyzing the distribution of hadrons, we may for example determine the fragmentation function $D_{f\rightarrow h}(z)$, i.e. the distribution of hadron species $h$ with an energy $E_h = zE_f$ originating from a shower initiating parton $f$ where $E_f$ is the whole energy of the jet.

\subsection{Spacetime structure of the shower}

While the vacuum shower evolution equations above are solved in momentum space only, the interaction with the medium requires modelling of the shower evolution in position space as well, because the medium properties in a general medium change as a function of the position space variables. Usually, these are given in the c.m. frame of the collision in terms of the spacetime rapidity $\eta_s$, the radius $r$, the proper time $\tau$ and the angle $\phi$, and knowledge of the medium evolution implies knowledge of medium properties such as the local medium temperature $T$ in the form $T(\eta_s, r, \phi,\tau)$.

In order to make the link from momentum space to momentum space, we assume that the average formation time of a shower parton with virtuality $Q$ is developed on the timescale $1/Q$, i.e. the average lifetime of a virtual parton with virtuality $Q_b$ coming from a parent parton with virtuality $Q_a$ is in the rest frame of the original hard collision (the local rest frame of the medium may be different by a flow boost as the medium may not be static) given by

\begin{equation}
\label{E-Lifetime}
\langle \tau_b \rangle = \frac{E_b}{Q_b^2} - \frac{E_b}{Q_a^2}.
\end{equation}  

Going beyond the ansatz of \cite{YAS,YAS2} where we used this average formation time for all partons, in the present work we assume that the actual formation time can be obtained from a probability distribution 

\begin{equation}
\label{E-RLifetime}
P(\tau_b) = \exp\left[- \frac{\tau_b}{\langle \tau_b \rangle}  \right]
\end{equation}

which we sample to determine the actual formation time of the fluctuation in each branching. This establishes the temporal structure of the shower. With regard to the spatial structure, we make the simplifying assumption that all partons probe the medium along the eikonal trajectory of the shower initiating parton, i.e. we neglect the small difference of the velocity of massive partons to the speed of light and possible (equally small) changes of medium properties within the spread of the shower partons transverse to its axis. 

\subsection{The parton-medium interaction}

In the following, we assume that any effect of the medium will affect the partonic stage of the evolution, but not the hadronization. This is equivalent to the idea that hadronization takes place outside the medium, an assumption commonly made also for leading parton energy loss calculations. The validity of this assumption will be dicussed below.

We use three different scenarios to model the interaction of partons with the medium. The first one, in the following referred to as RAD, has been used previously in \cite{YAS,YAS2}. The relevant property of the medium probed is the transport coefficient $\hat{q}(\eta_s, r ,\phi,\tau)$ which represents the virtuality gain $\Delta Q^2$ per unit pathlength of a parton traversing the medium. Note that this represents an average transfer, i.e.\ a picture which would be realized in a medium which is characterized by multiple soft scatterings with the hard parton. However, unlike in \cite{YAS,YAS2} the virtuality transfer to a shower parton is randomized in the present work since the formation time is distributed randomly around its average. Thus, effectively the present scenario includes the possibility to have both a small formation time and hence a small virtuality gain and a large formation time corresponding to a more substantial increase in virtuality. 

In practice, we increase the virtuality of a shower parton $a$ propagating through a medium with specified $\hat{q}(\eta_s, r ,\phi,\tau)$ by

\begin{equation}
\label{E-Qgain}
\Delta Q_a^2 = \int_{\tau_a^0}^{\tau_a^0 + \tau_a} d\zeta \hat{q}(\zeta)
\end{equation}

where the time $\tau_a$ is given by Eq.~(\ref{E-RLifetime}), the time $\tau_a^0$ is known in the simulation as the endpoint of the previous branching process and the integration $d\zeta$ is along the eikonal trajectory of the shower-initiating parton. If the parton is a gluon, the virtuality transfer from the medium is increased by the ratio of their Casimir color factors, $3/\frac{4}{3} =  2.25$.

If $\Delta Q_a^2 \ll Q_a^2$, holds, i.e. the virtuality picked up from the medium is a correction to the initial parton virtuality, we may add $\Delta Q_a^2$ to the virtuality of parton $a$ before using Eq.~(\ref{E-Qsq}) to determine the kinematics of the next branching. If the condition is not fulfilled, the lifetime is determined by $Q_a^2 + \Delta Q_a^2$ and may be significantly shortened by virtuality picked up from the medium. In this case we iterate Eqs.~(\ref{E-Lifetime}),(\ref{E-Qgain}) to determine a self-consistent pair of $(\langle \tau_a\rangle, \Delta Q^2_a)$. This ensures that on the level of averages, the lifetime is treated consistently with the virtuality picked up from the medium. The actual lifetime is still determined by Eq.~(\ref{E-RLifetime}).

In a second scenario, in the following called DRAG, we assume that the medium exerts a drag force on each propagating parton. The medium is thus characterized by a drag coefficient $D(\eta_s, r ,\phi,\tau)$ which describes the energy loss per unit pathlength. 

In the simulation, the energy (and momentum) are reduced by

\begin{equation}
\label{E-Drag}
\Delta E_a = \int_{\tau_a^0}^{\tau_a^0 + \tau_a} d\zeta D(\zeta)
\end{equation}

Again, for a gluon the energy loss is increased by the color factor ratio 2.25. As in the previous case, the energy loss induced by the drag force is randomized even given the branching kinematics due to the randomized formation time of a branching.

The third scenario has been suggested in \cite{JEWEL,HBP}. In the following, it is referred to as FMED. Here, the modification does not concern the parton kinematics, but rather the evolution kernel, Eqs.~(\ref{E-qqg}--\ref{E-gqq}). In this scenario, the singular part of the branching kernel in the medium is enhanced by a factor $1+f_{med}$, e.g. Eq.~(\ref{E-qqg}) becomes in the medium

\begin{equation}
P_{q\rightarrow qg}(z) = \frac{4}{3} \frac{1+z^2}{1-z} \Rightarrow \frac{4}{3} \left( \frac{2 (1+f_{med})}{1-z} - (1+z)\right)
\end{equation}

The effect of the medium is thus summarized in the value of $f_{med}$. Note that in the FMED scenario, no explicit reference to the spacetime structure of the shower is made, in this sense, the scenario is rather different from the other two.

Note that in the RAD scenario the shower {\em gains} energy from the medium by means of the virtuality increase, in the DRAG scenario the shower {\em loses} energy to the medium whereas the shower energy is conserved in the FMED scenario. While this appears surprising at first, it is actually rather a matter of book-keeping. For a shower in the medium, there is no conceptual way to separate soft partons from the shower and from the medium. However, the model framework outlined above does not treat the medium as consisting of partons, but rather as an effective influence on the shower. Thus, in a more realistic model one would define a criterion (say a momentum scale) based on which partons are removed from the shower and become part of the medium. In such a model, all three scenarios would lead to a loss of energy from the shower to the medium through the appearance of soft partons in the evolution, in addition to possible other mechanisms of energy transfer to the medium.

\section{Comparison for a constant medium}

In this section, we perform several computations for the simple case of a constant medium with fixed pathlength. This is chiefly done in order to establish the relation of the models outlined above to older computations based on leading parton energy loss.

\subsection{Presence and absence of scaling}

\label{S-Scaling}

\begin{figure*}[htb]
\begin{center}
\epsfig{file=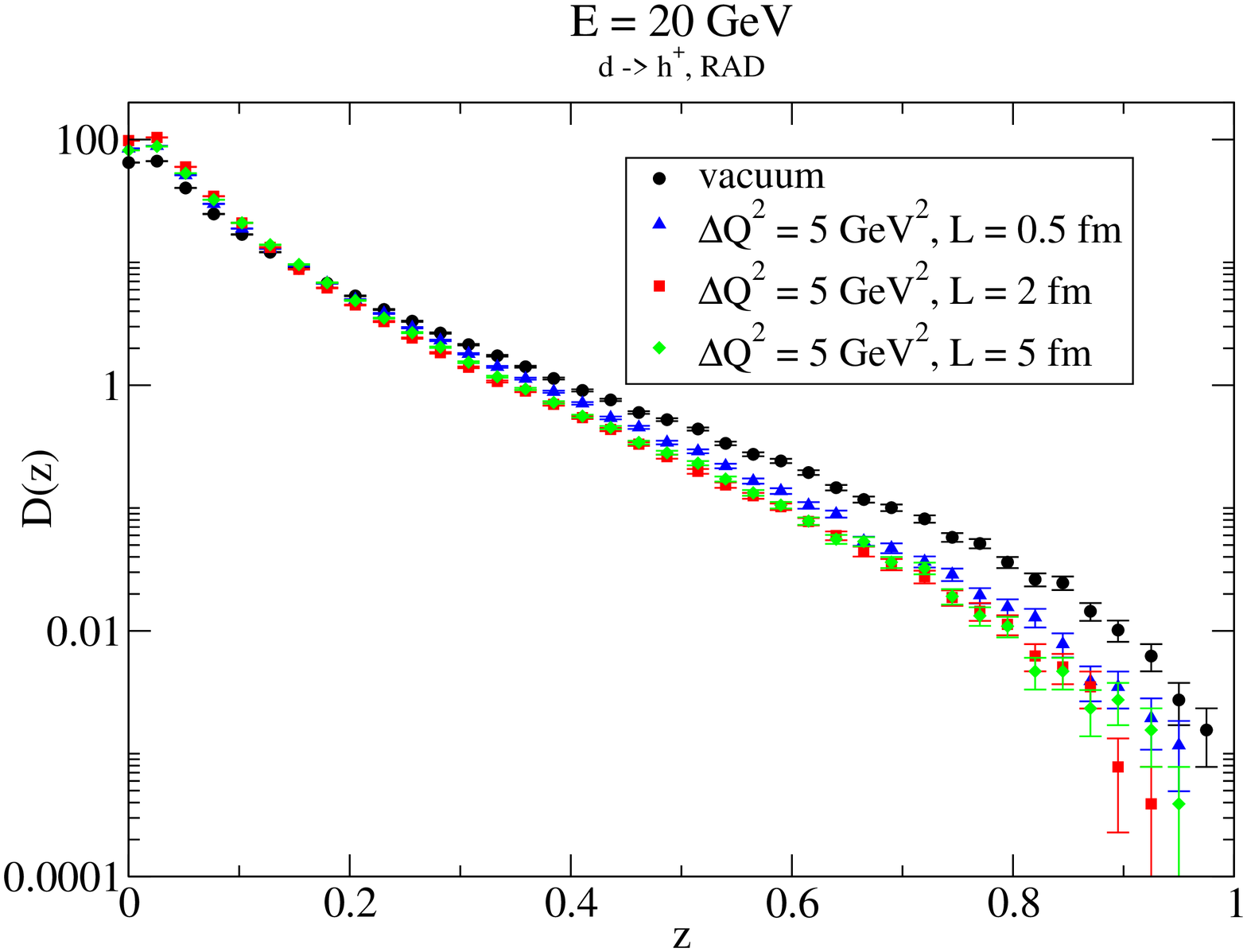, width=8cm}\epsfig{file=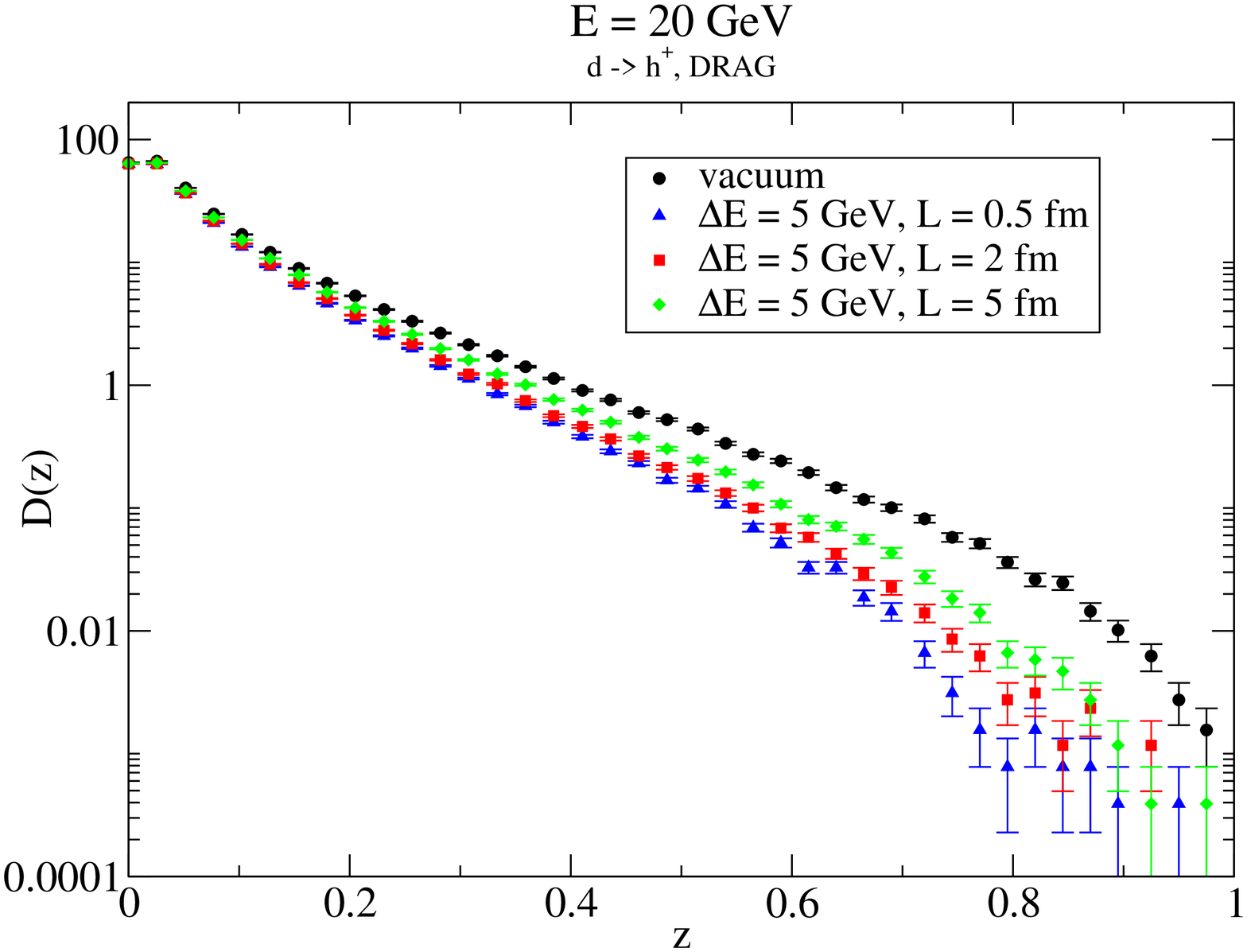, width=8cm}
\end{center}
\caption{\label{F-Scaling1}The MMFF of a $d$-quark into charged hadrons for constant value of $\Delta Q^2_{tot} = \hat{q}L = 5$ GeV$^2$ in the RAD scenario (left panel) and $\Delta E_{tot} = DL = 5$ GeV in the DRAG scenario (right panel) for different pathlengths in a constant medium. }
\end{figure*}

A constant medium corresponds to a choice of a single value of $\hat{q}, D$ or $f_{med}$. However, in the case of both the RAD and the DRAG scenario, also the medium length $L$ has to be specified, thus in principle the medium is characterized by two parameters. In \cite{YAS} however we found an approximate scaling law for the RAD scenario according to which the modification chiefly depends on the virtuality picked up along the eikonal path of the shower initiating parton $\Delta Q^2_{tot} = \int d \zeta \hat{q}(\zeta)$ or in the case of a constant medium simply $\hat{q} L$. A similar scaling law can also be established for the DRAG scenario, albeit only in the case of an expanding medium (see below). Whenever such a scaling law holds, a comparison between the different scenarios can be made based on the single parameter $\Delta Q^2_{tot}$ or $\Delta E_{tot}$ only.

It is clear that the scaling cannot work for all the possible functional forms $\hat{q}(\zeta)$. In two different limits this can be made plausible: If, in the RAD scenario, $\Delta Q^2_{tot}$ is added at once initially, $\Delta Q^2/Q^2$ is for reasonable values of hard process kinematics and medium properties very small. For example, for typical RHIC kinematics the initial $Q^2$ from which the evolution starts may be 400 GeV$^2$ whereas the total virtuality acquired for a parton traversing the whole medium is about 15 GeV$^2$ according to the results of \cite{YAS}. However, such a small correction will not influence the shower evolution significantly. On the other hand, if the virtuality is added later when the typical $Q^2$ is of order of $\Delta Q^2$, a much stronger modification is expected. Thus, one expects the scaling law to be violated into the direction of less medium effect if $\hat{q}(\zeta)$ is strongly peaked towards $\tau=0$. 

A similar argument, can be made for the DRAG scenario. The drag force acts on every parton in the shower. This means that if $D(\zeta)$ is strongly peaked towards $\tau=0$, then the drag force acts only on one parton, the shower initiator, whereas if it is applied later, its effect is felt by several partons. 

On the other hand, note that the shower evolution is terminated for every parton which reaches $Q^2 \le Q^2_{min} =1$ GeV$^2$. This implies that the typical lifetime of the shower for an initial parton with energy $E$ is given by $\tau_{max} \sim E/Q^2_{min}$, thus a shower with $E=20$ GeV probes the medium on average for a distance of 4 fm (Eq.~(\ref{E-RLifetime}) leads to fluctuations around this average though). Thus, if $L$ is chosen much beyond $\tau_{max}$, $\hat{q}L$ or $DL$ are not good parameters any more, as the shower does not effectively probe the whole medium.

The latter effect is clearly not related to an actual physics effect but rather an artefact of the need to switch to a non-perturbative description of hadronization at some point in the simulation. It is unreasonable that a parton (or proto-hadron) would feel no effect from the medium just because its virtuality is small, however it is unclear just how the effect should be implemented properly in the present framework. The behaviour of the simulation thus depends on the actual choice of $Q_{min}$, and this needs to be optimized eventually in comparison with data. A study of the effect of changing $Q_{min}$ will be presented below.

The resulting MMFF for a light quark into charged hadrons for constant $\hat{q}L$ or $DL$ and a variation of pathlength is shown in Fig.~\ref{F-Scaling1} for both the RAD and the DRAG scenario. In a constant medium, the RAD scenario shows approximate scaling for pathlength between 0.5 and 5 fm. The DRAG scenario does not exhibit a strong scaling in the region of large $z$, but in the region $z \sim 0.5$ which is predominantly probed when computing the single hadron spectra, the variations are not too large for pathlengths between 0.5 and 3 fm.

Note that neither the short pathlength nor the long pathlength limit is actually problematic for a realistic medium evolution taken from a hydrodynamical model. The first limit is avoided by virtue of the thermalization time of order $O(0.6)$ fm for RHIC kinematics. This is a large time compared with the timescale in which the first branchings in a shower occur, thus by the time the medium is present, the shower is already well developed. The second limit is avoided because in an expanding medium $\hat{q}(\zeta)$ or $D(\zeta)$ drop rapidly as a function of time, thus the late time contribution to $\int d \zeta \hat{q}(\zeta)$ or $\int d \zeta D(\zeta)$ is small in any case. Thus, the scaling works much better for a realistic evolution as the constant medium results would suggest.

\subsection{Energy loss and quenching weights}

\begin{figure*}[htb]
\begin{center}
\epsfig{file=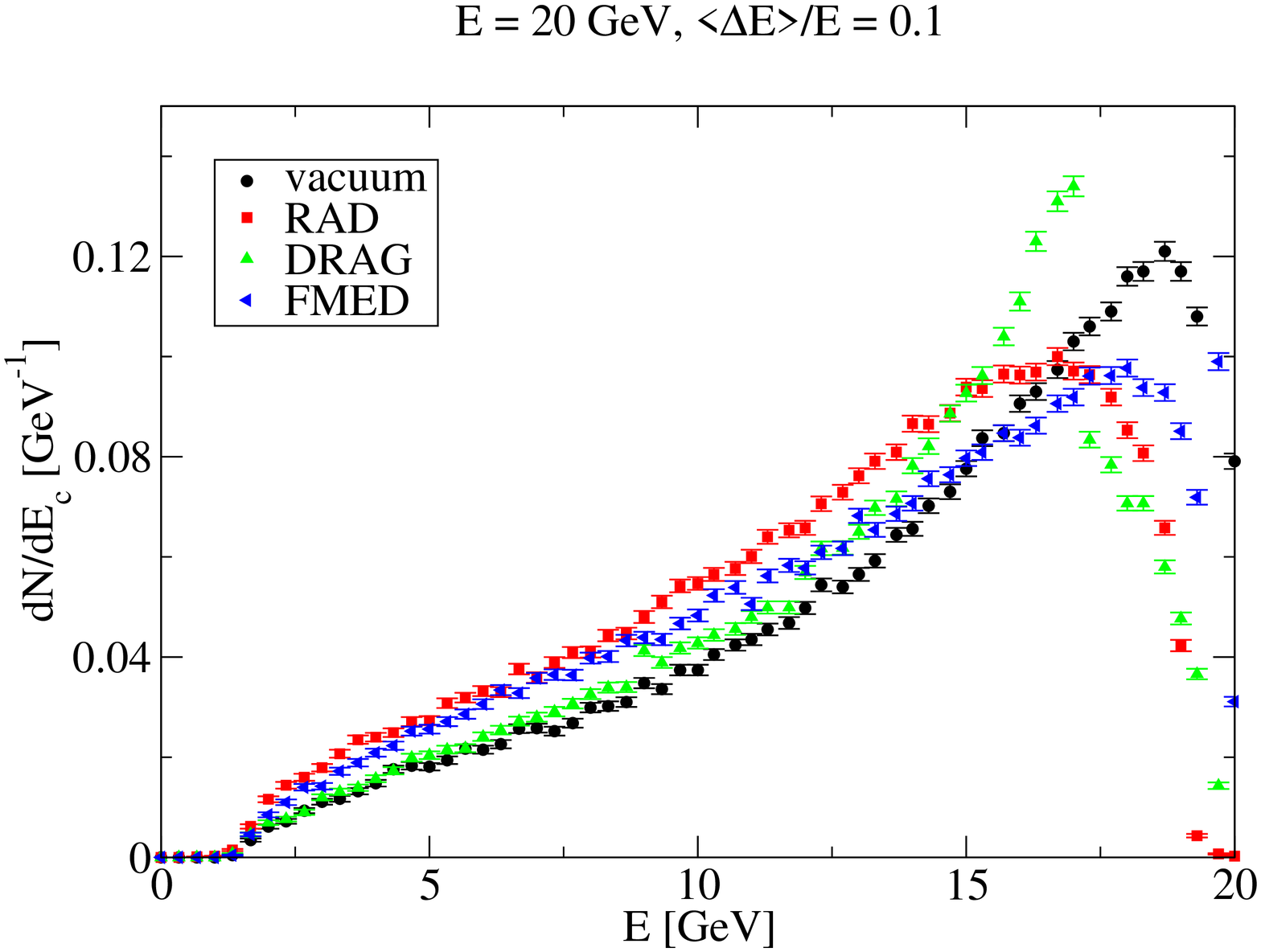, width=8cm}\epsfig{file=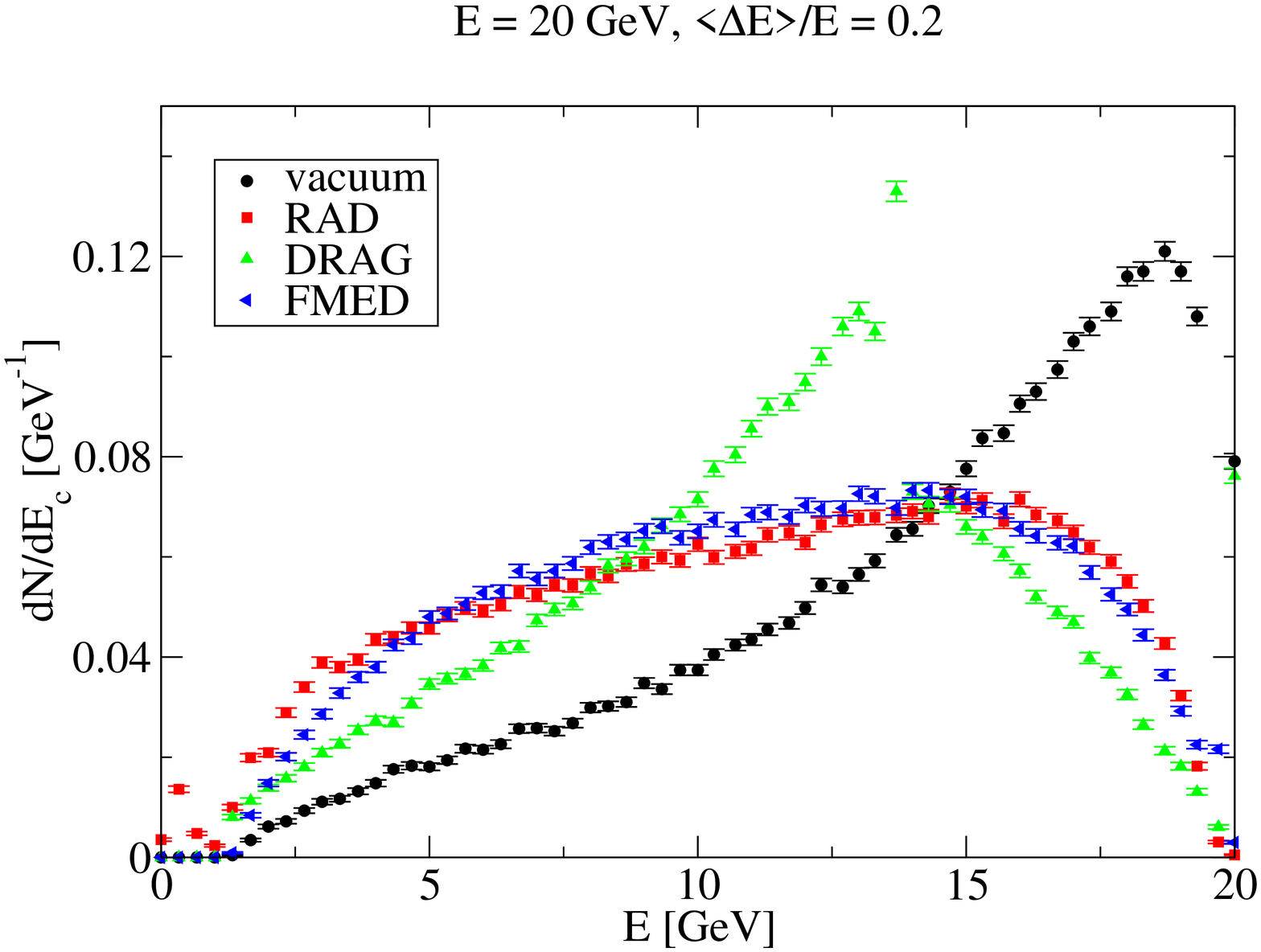, width=8cm}
\end{center}
\caption{\label{F-cdist}Energy distribution of the leading $c$ quark for a 20 GeV $c$ quark as shower initiator in the three different scenarios for the parton-medium interaction (see text). Left panel: 10\% average energy loss, right panel: 20\% average energy loss.}
\end{figure*}

We now proceed to compare the three scenarios on the basis of leading parton energy loss. This is relevant to make the connection to previous calculations in the BDMPS or ASW formalism \cite{Jet2,QuenchingWeights} which are formulated using this concept. For this purpose, we select the shower initator to be a $c$-quark and extract the energy distribution of the leading $c$-quark $dN/dE_c$ after the shower. From the comparison of the distribution $dN/dE_c^{vac}$ in vacuum and in the medium $dN/dE_c^{med}$, we can deduce the energy loss probability distribution $P(\Delta E)$. The idea is to make an ansatz 

\begin{equation}
\label{E-Folding}
\frac{dN}{dE}_c^{med}\negthickspace \negthickspace(E) = \int d (\Delta E) \frac{dN}{dE}_c^{vac}\negthickspace\negthickspace(E') P(\Delta E) \delta(E' - E - \Delta E)
\end{equation}

and solve it for $P(\Delta E)$. Note that this ansatz contains the rather drastic assumption that there is no parametric dependence on the initial energy $E$. If we require $P(\Delta E)$ to be a probability distribution, the assumption may imply that for some partons in the distribution $dN/dE_c^{vac}$ the energy loss $\Delta E$ is larger than their energy $E$ in which case they have to be considered lost to the medium. A similar situation also occurs in the application of the ASW formalism to finite energy kinematics. The problem of the validity of assuming energy independence however only concern the comparison with the ASW results in which energy loss is formulated in terms of a probability density $P(\Delta E)$. In all other results presented in this manuscript, the full information of the shower including finite energy kinematics is used and no assumption about energy independence of energy loss needs to be made.

The choice of a $c$-quark as shower initiator has a twofold motivation. First, it allows to define energy loss in the same way as done in the ASW formalism. Note that the ASW formalism assumes infinite parent parton energy and calculates energy loss via the radiation spectrum off the parent. In applying the formalism to finite energy, a process may occur in which a radiated gluon takes 90\% of the energy of an initial quark $q_1$. This energy is then considered to be lost from the $q_1$. However, in the shower language, the radiated gluon would in this case become the {\em new} leading parton, and even tagging the leading quark out of a shower would not prevent processes where this gluon splits into a $q\overline q$ pair where the new quark $q_2$ might still be harder as the original parent $q_1$ of the gluon. The choice of a $c$ quark as shower initiator effectively suppresses such processes and allows to treat energy loss as closely as possible to ASW \cite{Urs}.

The second advantage of extracting $P(\Delta E)$ from a $c$ quark is that the $c$-fragmentation is rather hard, i.e. the probability distribution to find the leading $c$-quark after a vacuum shower peaks close to $z=1$. This effectively means that if one considers an additional, medium-induced shift in energy, most of the energy range is still available for the dominant part of the distribution. This is very different for a light quark shower where the leading quark distribution typically peaks at $z \sim 0.5$ and any energy loss of $\Delta E > E/2$ shifts the bulk of the distribution into the unphysical region of negative energies.

In Fig.~\ref{F-cdist} we show the leading charm distributions both in vacuum and in medium for a medium pathlength of $L=2$ fm. In order to make a meaningful comparison between the different scenarios, the average relative energy loss $\langle \Delta E \rangle/E$ is fixed to 10\% or 20\% respectively. 

\begin{figure*}[htb]
\begin{center}
\epsfig{file=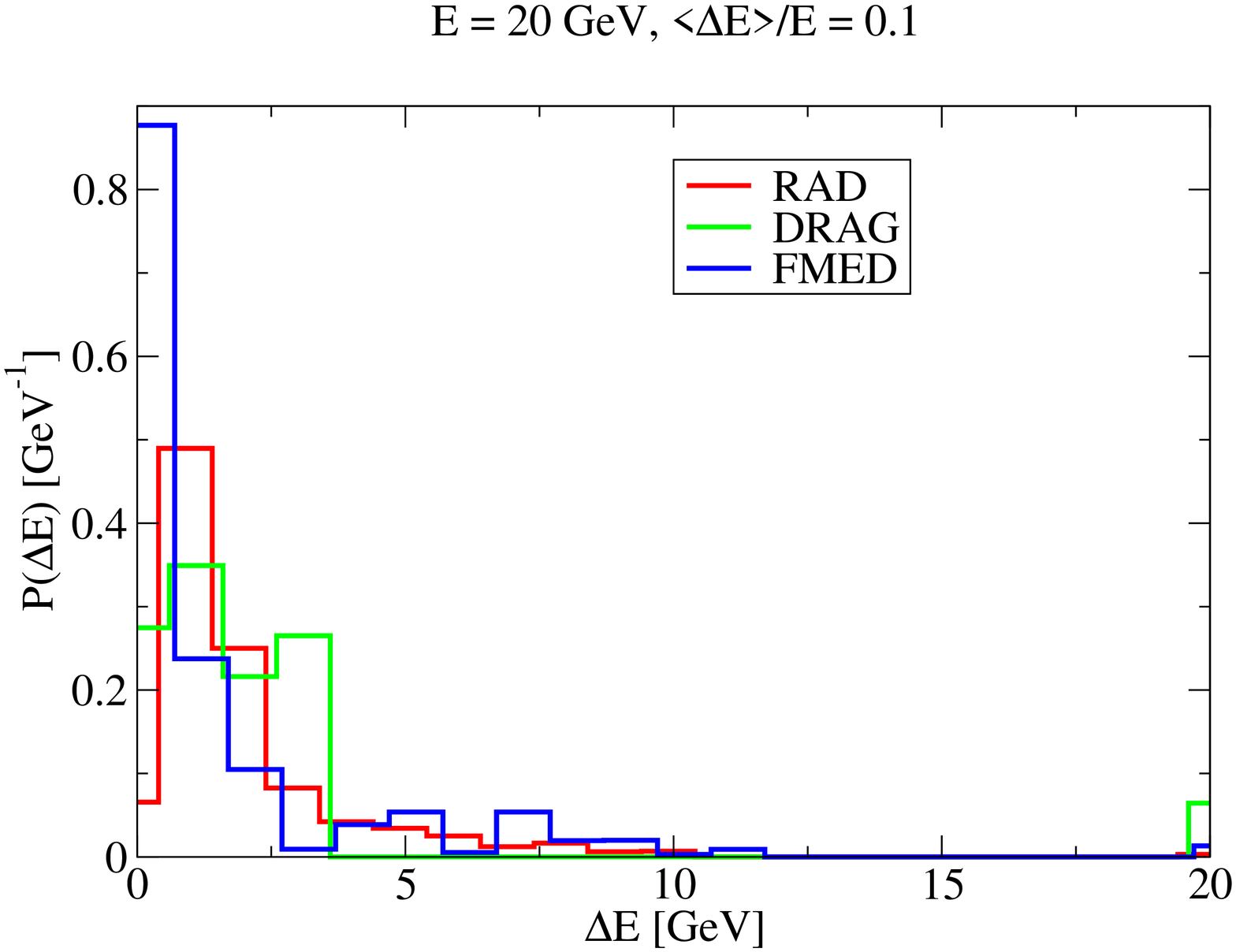, width=8cm}\epsfig{file=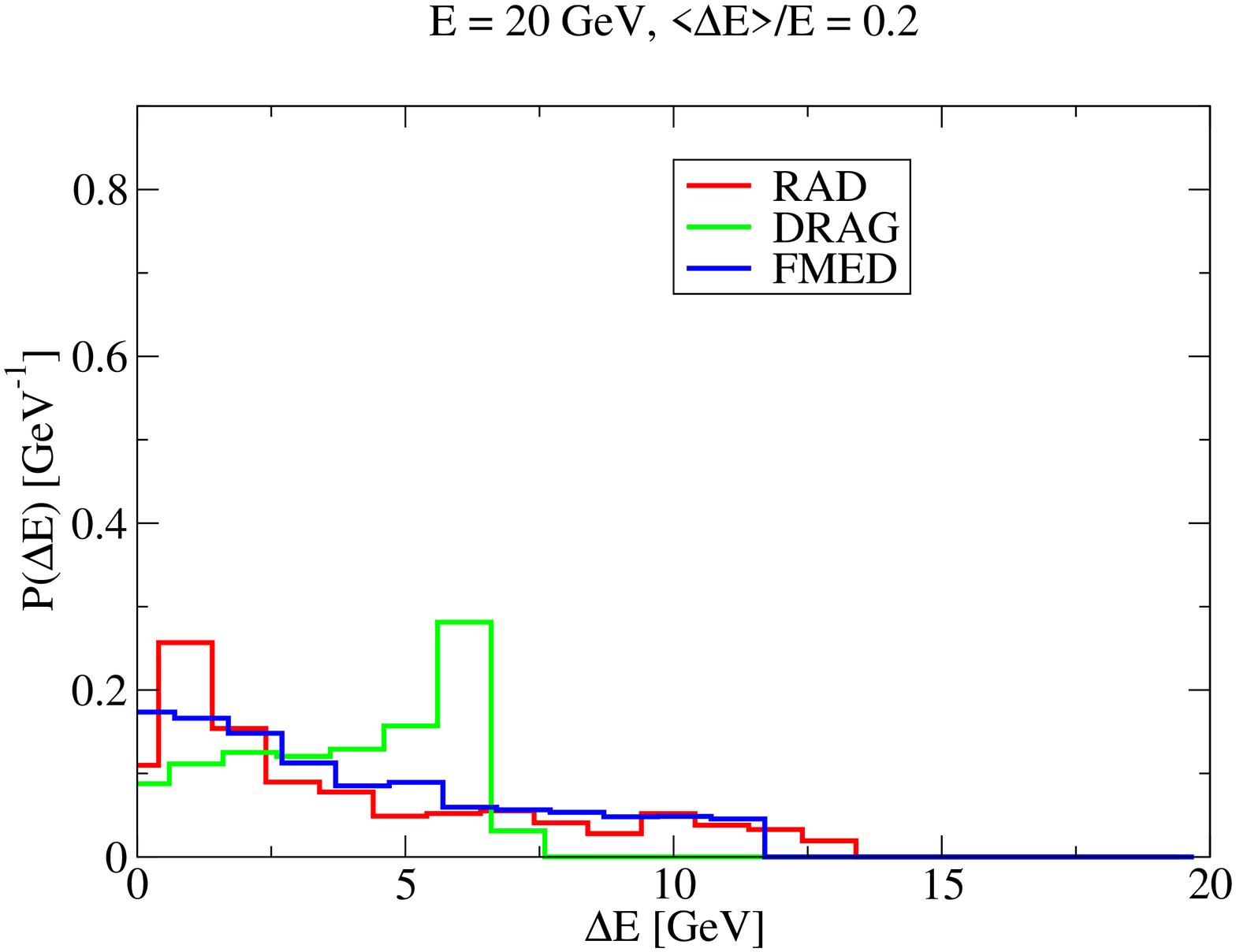, width=8cm}
\end{center}
\caption{\label{F-PDeltaE}Energy loss probability distribution $P(\Delta E)$ for the leading $c$-quark for a 20 GeV $c$-quark as shower initiator in the three different scenarios for the parton-medium interaction (see text). Left panel: 10\% average energy loss, right panel: 20\% average energy loss.}
\end{figure*}

In order to deduce the energy loss probability distribution from these results, we have to solve Eq.~(\ref{E-Folding}). By discretizing the integral over $\Delta E$ in Eq.~(\ref{E-Folding}) we can cast it into the form of a matrix equation

\begin{equation}
\label{E-Matrix}
N_i(E^i) = \sum_{j=1}^n K_{ij}(E^i, \Delta E^j) P_j(\Delta E^j)
\end{equation}

where $dN/dE_c$ is provided at $m$ discrete values of $E$ labelled $N_i$ and $P(\Delta E)$ is probed at $n$ discrete values of $\Delta E$ labelled $P_j$. The kernel $K_{ij}$ is then the calculated $dN/dE_c$ for all pairs $(E^i, \Delta E^j)$ where the energy loss acts as a shift of the distribution, i.e. $dN/dE_c^{med}(E) = dN/dE_c^{vac}(E+\Delta E)$.

Eq.~(\ref{E-Matrix}) can in principle be solved for the vector $P_j$ by inversion of $K_{ij}$ for $m=n$. However, in general this does not guarantee that the result is a probability distribution. Especially in the face of statistical errors and finite numerical accuracy the direct matrix inversion may permit negative $P_j$ which have no probabilistic interpretation.

Thus, a more promising solution which avoids the above problems is to let $m > n$ and find the vector $P$ which minimizes $|| N - K P||^2$ subject to the constraints $0 \le P_i \le 1$ and $\sum_{i=1}^n P_i = 1$. This guarantees that the outcome can be interpreted as a probability distribution and since the system of equations is overdetermined for $m>n$ errors on individual points $R_i$ do not have a critical influence on the outcome any more. This is the approach we have chosen.

The results are shown for $L=2$ fm in Fig.~\ref{F-PDeltaE}. Qualitatively, both the radiative energyloss scenarios RAD and
FMED produce energy loss probability distributions which are similar to the ASW quenching weights \cite{QuenchingWeights} in the sense that they are flat across a wide range in $\Delta E$. In contrast, the DRAG scenario produces a localized peak in the energy loss distribution which reminds of the quenching weights found for elastic energy loss scenarios \cite{Wicks,Elastic}. Especially for larger energy loss, the RAD and the FMED scenario lead to almost identical results.

However, there is an important difference to the ASW quenching weights: While the ASW results typically show a large discrete probability for no energy loss, the results obtained here show no substantial strength in the first bin (the inversion procedure outlined above cannot separate zero energy loss from small energy loss).

\subsection{Parametric dependence of mean energy loss}

\begin{figure}[htb]
\epsfig{file=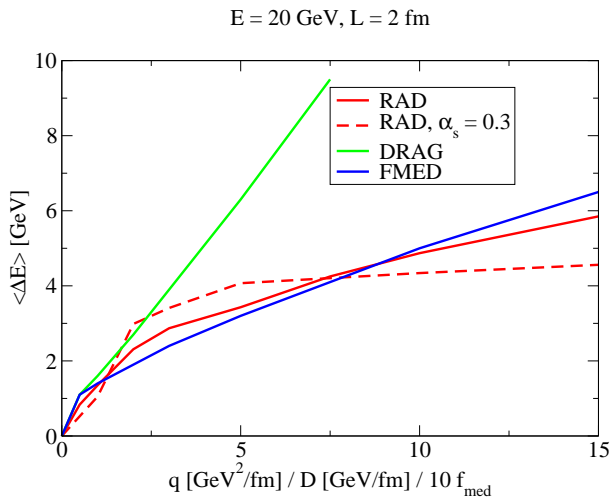, width=8cm}
\caption{\label{F-DE}Mean energy loss as a function of the medium properties in different scenarios for the parton-medium interaction in a constant medium with $L=2$ fm.}
\end{figure}

In order to gain more insight into the different scenarios, we investigate in Fig.~\ref{F-DE} for a constant medium with $L=2$ fm how the mean energy loss, defined as $\langle \Delta E \rangle = \int d \Delta E \Delta E P(\Delta E)$ with $P(\Delta E)$ obtained as in the previous section behaves as a function of the relevant medium parameters. We include a scenario in which the strong coupling constant is not allowed to run with the virtuality scale in the shower (as is the default option in PYSHOW) but is kept fixed at $\alpha_s = 0.3$.

There is no unique way to present and compare the results, as the three relevant parameters $\hat{q}, D$ and $f_{med}$ are rather different. However, as apparent from Fig.~\ref{F-DE}, it is possible to find a simple proportionality relation between $\hat{q}$ and $f_{med}$ such that the rise of the mean energy loss appears very similar. This, in addition to the similarity of $P(\Delta E)$ for both the RAD and the FMED scenario points towards some generic properties of radiative energy loss scenarios independent of the details of the implementation.

In particular, the RAD and the FMED scenario exhibit saturation of the mean induced energy loss at about 25\% of the total energy as the medium effect is increased. This saturation is even more pronounced for a constent $\alpha_s$. In striking contrast, the DRAG scenario in which energy is directly transferred to the medium shows an almost linear rise up to mean energy losses of 50\%. Note that the extraction of the energy loss probability based on discretization and matrix inversion as outlined above becomes increasingly problematic at $\langle \Delta E \rangle /E  > 0.4$ due to the problem of partons being shifted to negative energies mentioned above.

\section{Comparison for a single path in an expanding medium}

We now turn to a more realistic scenario in which the parton propagates in a  medium as created in a heavy-ion collision. Relativistic fluid-dynamical models such as \cite{Hydro3d} give a good description of many bulk properties of the medium, hence in the following we will assume that hydrodynamics is a valid description of the medium. Both the finite size and the finite lifetime of such a medium are felt by the parton. In particular, the local density may drop a) because of a spatial variation, i.e. the parton reaches the medium edge and b) a temporal variation, i.e. the global expansion of the medium reduces the overall density as a function of time. In addition, there are arguments that the hydrodynamical flow of the medium expansion should also have a direct influence on the medium properties as seen by the medium due to Lorentz transformation between the moving local medium rest frame and the frame of the hard collision \cite{Flow1,Flow2}.

\subsection{Characterization of the medium}

In \cite{YAS} we have established that if $\hat{q}$ is linked with the medium properties by the relation

\begin{equation}
\label{E-qhat}
\hat{q}(\zeta) = K \cdot 2 \cdot [\epsilon(\zeta)]^{3/4} (\cosh \rho(\zeta) - \sinh \rho(\zeta) \cos\psi)
\end{equation}

with $K$ a parameter determining the interaction strength which is {\em a priori} unknown (in an ideal QGP, $K=1$ is expected \cite{Baier} but a comparison study of different energy loss models has shown to be inconclusive in extracting values for $K$ \cite{HydroComp}), the medium energy density $\epsilon$, the local flow rapidity $\rho$ with angle $\psi$ between flow and parton trajectory \cite{Flow1,Flow2}, we find that the vast majority of paths found in the 3-dimensional hydrodynamical model of Bass and Nonaka \cite{Hydro3d} leads to a $\hat{q}(\zeta)$ which can be described by the rather simple expression

\begin{equation}
\hat{q}(\zeta) = \frac{a}{(b+\tau/(1 fm/c))^c}.
\end{equation}

Based on this expression, we investigated three different scenarios (approximately representing a parton travelling into $+x$ direction originating from $x=4$ fm (A), $x=0$ (B) and $x=-4$ fm (C), $y=0$ in all cases in the transverse $(x,y)$ plane at midrapidity. These trajectories are characterized by the parameters $(b=1.5, c=3.3, \tau_E= 5.8$ fm/c$)$ (A),  $(b=1.5, c=2.2, \tau_E= 10$ fm/c$)$ (B) and $(b=1.5, c=2.2, \tau_E= 15$ fm/c$)$ (C) and are quite typical for partons close to the surface (A), emerging from the central region (B) or traversing the whole medium (C). As in \ref{S-Scaling} in the present paper for a constant medium,  we found that an approximate scaling in which the medium effects did not depend on details of the trajectories (A), (B), or (C) but only on $\Delta Q^2_{tot} = \int d \zeta \hat{q}(\zeta)$.

The virtue of this scaling law is twofold: First, it allows to present the medium modifications for the relevant class of functions $\hat{q}(\zeta)$ as a function of a single parameter $\Delta Q^2_{tot}$ only. Second, it considerably speeds up the computation for a comparison with data where a weighted average over {\em all} possible paths through the medium has to be computed. 

In \cite{YAS} we have made the rather drastic assumption that the medium does not exert any effect before the thermalization of the medium at the time $\tau_{in}$ where $\tau_{in} = 0.6$ fm/c in the model studied for RHIC \cite{Hydro3d}. In the following, we adopt a more realistic approach in which we increase the medium effect linearly from zero at $\tau=0$ to its value reached at $\tau_{in}$. The idea behind this is that initially no medium can be present, as the timescale for hard processes precedes any other timescale in the system. However, even a medium which is not yet equilibrated may interact with hard partons and lead to scattering processes. A linear interpolation between the initial time and the equilibration time seems a reasonable prescription to capture part of these effects. In practice, qualitative aspects of the results of \cite{YAS}, in particularly the presence of the scaling, are not substantially altered by this modification. There is however an effect on the numerical value of extracted medium parameters. 

Let us now consider the other scenarios DRAG and FMED. Eq.~(\ref{E-qhat}) which links $\hat{q}$ with the hydrodynamical properties of the medium is based on counting the potential scattering centers along the parton trajectory. $\epsilon^{3/4}$ for an ideal gas corresponds to the entropy density, which in turn is proportional to the medium density. The additional factor $(\cosh \rho(\zeta) - \sinh \rho(\zeta) \cos\psi)$ is nothing but the appropriate transformation to determine how the density seen by the parton is changed under a boost of the restframe of the medium \cite{Flow1}. It is reasonable to assume a similar measure of potential scattering centers to be relevant for the other scenarios. This ansatz leads to 

\begin{equation}
\label{E-D}
D(\zeta) = K_D \cdot [\epsilon(\zeta)]^{3/4} (\cosh \rho(\zeta) - \sinh \rho(\zeta) \cos\psi)
\end{equation}

for the drag coefficient $D$ with an {\em a priori} unknown parameter $K_D$ specifying the overall strength of the drag force.

As discussed above, the FMED scenario has no explicit dependence on the spacetime evolution of the shower, but it seems reasonable the the parameter $f_{med}$ should depend on the total effect of the medium measured in the number of potential scatterers which have been encountered. This leads to the ansatz

\begin{equation}
\label{E-fmed}
f_{med} = K_f \int d \zeta [\epsilon(\zeta)]^{3/4} (\cosh \rho(\zeta) - \sinh \rho(\zeta) \cos\psi).
\end{equation}

Here, as in the previous scenarios, we also introduce an {\em a priori} unknown parameter $K_f$ which determines the strength of the parton-medium interaction.

\begin{figure}[htb]
\epsfig{file=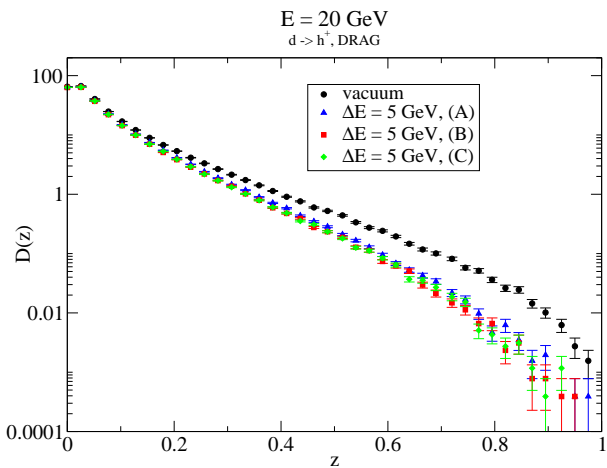, width=8cm}
\caption{\label{F-ScalingD1}The MMFF of a 20 GeV $d$-quark into charged hadrons for three different paths (A), (B) and (C) (see text) with $\Delta E_{tot} = 5$ GeV in the DRAG scenario. }
\end{figure}

The scaling within the RAD scenario of the results with $\Delta Q^2_{tot}$ has been established in \cite{YAS} and in the present paper also for a constant medium in \ref{S-Scaling}. The DRAG scenario shows no strong scaling for a constant medium, but as anticipated the result is more promising for an expanding medium. The validity of the scaling under these conditions is apprent from Fig.~\ref{F-ScalingD1} where we compute for fixed $\Delta E$ for the three different paths (A), (B) and (C). Note that scaling of the FMED scenario is realized by definition using the ansatz Eq.~(\ref{E-fmed}).

\subsection{Longitudinal momentum distribution of the shower}

\begin{figure*}[htb]
\begin{center}
\epsfig{file=D_comp.eps, width=8cm}\raisebox{-2mm}{\epsfig{file=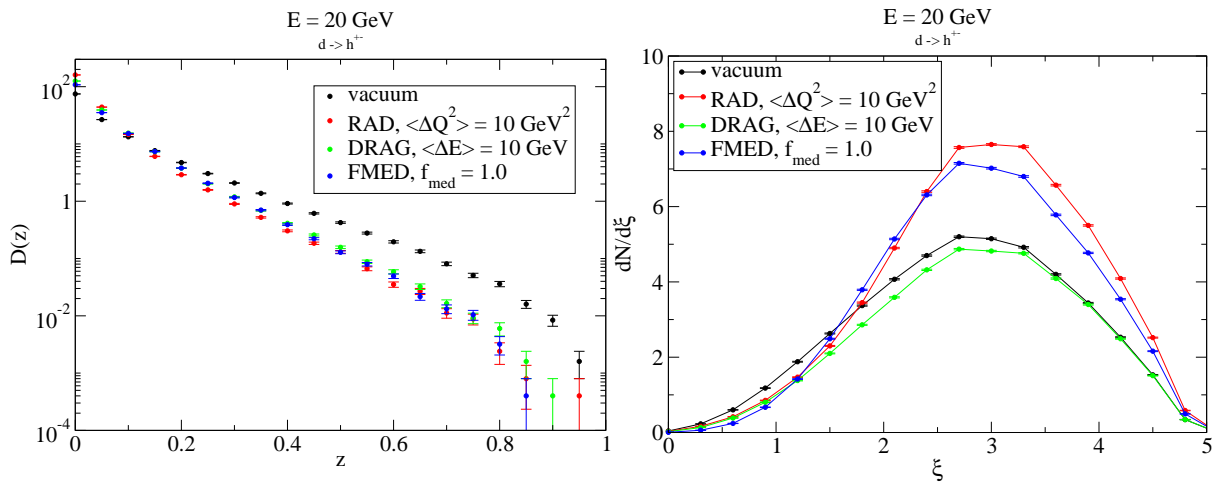, width=8cm}}
\end{center}
\caption{\label{F-LComp}Longitudinal momentum distribution of charged hadrons inside a jet originating from a 20 GeV $d$-quark shown as fragmentation function $D(z)$ (left panel) and $dN/d\xi$ (right panel) for the vacuum and the three different scenarios for the parton-medium interaction (see text). The medium parameters have been chosen to let the modified $D(z)$ approximately agree for $0.4 < z < 0.7$ for all in-medium scenarios.}
\end{figure*}

In Fig.~\ref{F-LComp}, we show the longitudinal momentum distribution of charged hadrons inside the shower in therms of the MMFF $D(z)$ for three different scenarios in comparison. There is no a priori criterion at which values of the three medium parameters $\Delta Q^2_{tot}$, $\Delta E_{tot}$ and $f_{med}$ the three different scenarios should be compared. For the comparison in terms of energy loss probability distributions done above we required a fixed value $\langle \Delta E \rangle/E$, but this is not a meaningful variable when one wants to compare on the basis of the whole parton shower instead of the leading parton kinematics only. Here, we chose the criterion that the MMFF approximately agree in an interval of $0.4 < z < 0.7$. This is the region of the fragmentation function which is predominantly probed when the fragmentation function is folded with a pQCD parton spectrum to compute single inclusive hadron production. The implication is that a computation with MMFFs agreeing in the above interval would yield approximately the same observable hadron spectra. This choice leads to the interesting and amusing numerical coincidence that if the parameters are given in powers of GeV, the relation $\Delta Q^2_{tot}/\text{GeV}^2 \approx \Delta E/\text{GeV} \approx 10 f_{med}$ holds.

We show the MMFF of a 20 GeV $d$-quark into charged hadrons for $\Delta Q^2 = 10$ GeV$^2$ (the parameters of the other scenarios adjusted correspondingly) in Fig.~\ref{F-LComp}, right panel. 

In order to focus more on the hadron production at low momenta, we introduce the variable $\xi = \ln(1/x)$ where $x = p/E_{jet}$ is the fraction of the jet momentum carried by a particular hadron and $E_{jet}$ is the total energy of the jet. The inclusive distribution $dN/d\xi$, the so-called Hump-backed plateau, is an important feature of QCD radiation \cite{Muller,Dokshitzer} and is in vacuum dominated by color coherence physics. 

In Fig.~\ref{F-LComp} (left panel) we show $dN/d\xi$ for the three different scenarios in comparison with the unmodified result. It is apparent from the figure that while the three scenarios agree in the high $z$ and consequently low $\xi$ region, they exhibit sizeable differences in the high $\xi$ region where  induced radiation is expected to contribute to soft hadron production. Here, both the radiative scenarios RAD and FMED show the expected enhancement of the distribution, but the DRAG scenario is strikingly different --- it falls below the vacuum result. However, this is hardly surprising, as in this scenario energy is taken away from the evolving shower and is hence not available for hadron production. 

While a measurement of $dN/d\xi$ would appear to be a promising means to distinguish between induced radiation and a drag force as the microscopic realization of energy loss, it has to be pointed out that there are two things which urge some caution. First, the Lund scheme used to model hadronization in the present framework assumes that hadronization takes place far outside the medium. If the energy of a hadron $h$  of mass $m_h$ is $E_h$, the spatial scale at which hadronization occurs can be estimated as $l_h \approx E_h/m_h^2$. For pions, this is not a problem throughout the kinematic range, but for kaons and protons the hadronization length is considerably shortened. Even a 10 GeV proton has only $l_h  \approx 2$ fm, thus heavy hadron production in the high $\xi$ region is not addressed adequately in the model, as one cannot safely assume hadronization takes place outside the medium where the Lund model is applicable. Nevertheless, since pions constitute the bulk of charged hadron production, the essential features of the model are expected to be robust.

The second issue concerns the effect of trigger bias. A series of experimental cuts has to be imposed on events in heavy-ion collisions to discriminate hadrons belonging to jets from the background of soft medium hadrons. However, strongly modified jets (for example those emerging from the medium center) are less likely to fall within the cuts than unmodified jets (such as those from the medium edge). As a result there is a trigger bias which suppresses events in which a modification of $dN/d\xi$ is visible. A calculation in the RAD scenario taking into account a realistic series of experimental cuts has been performed in \cite{YAS2} and found that there should be no visible enhancement if jets are identified directly via a standard set of cuts.

\subsection{Angular distribution}

\begin{figure}[htb]
\epsfig{file=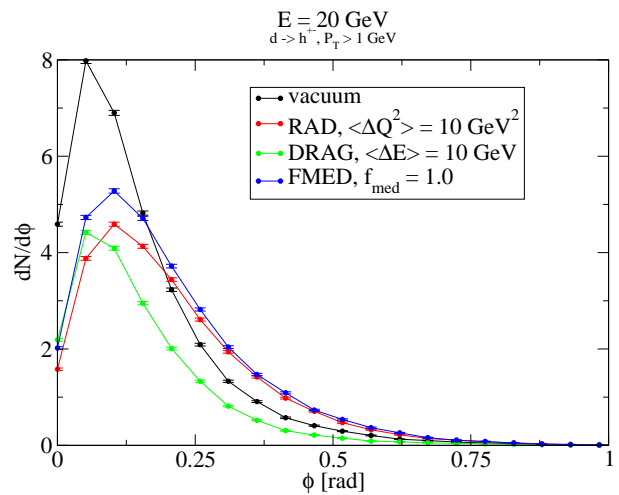, width=8cm}
\caption{\label{F-PhiComp}Angular distribution of charged hadrons above 1 GeV coming from the fragmentation  of a 20 GeV $d$-quark for vacuum and  three different scenarios of parton-medium interaction (see text).}
\end{figure}

Another possibility to identify the mechanism of the parton-medium interaction is to study the structure of the jet transverse to the jet axis. This is reflected e.g. in the angular distribution of hadrons around the jet axis. The distribution $dN/d\phi$ where $\phi$ is the angle between hadron and jet axis for $\Delta Q^2_{tot} = 10$ GeV$^2$ (the parameters in the other scenarios adjusted accordingly) for the vacuum and the three different scenarios is shown in Fig.~\ref{F-PhiComp} where a cut in momentum of 1 GeV has been applied to focus on hadrons which would appear above the soft background of a heavy-ion collision.

It is apparent from the figure that the radiative energy loss scenarios again roughly agree with each other and lead to angular broadening of the jet as compared to the vacuum result, whereas the DRAG scenario shows no indication for broadening. 

\subsection{The sensitivity to $Q_{min}$}

\label{S-Qmin}

For a constant medium, we noted earlier that there is a sensitivity to the choice of the minimum virtuality scale $Q_{min}$ at which partons in the shower are evolved further. Before comparing the results of this section to data, it is reasonable to ask to what extent a choice of $Q_{min}$ different from its default value $Q_{min}= 1$ GeV has an influence on the results.

\begin{figure}[htb]
\epsfig{file=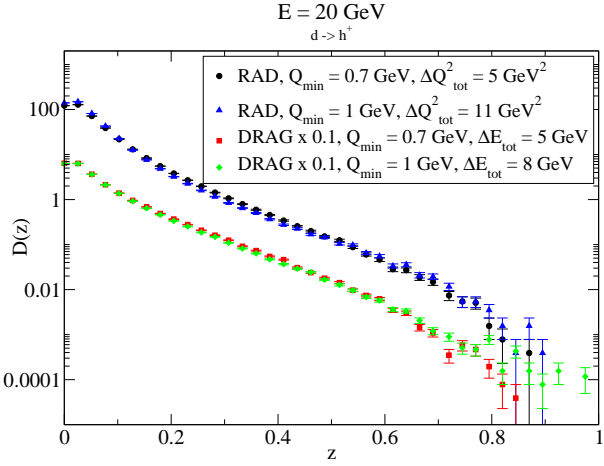, width=8cm}
\caption{\label{F-QComp}MMFF for a 20 GeV $d$-quark as shower initiator for different values of the minimum shower virtuality $Q_{min}$ in the RAD and the DRAG scenario (see text) where medium parameters have been adjusted to compensate for the choice of $Q_{min}$.}
\end{figure}

In Fig.~\ref{F-QComp} we show results for the MMFF in both the RAD and the DRAG scenario with a lower $Q_{min} = 0.7$ GeV where the shower evolves on average a factor two longer. Superimposed are results with the default choice $Q_{min} = 1$ GeV for which the medium parameters $\Delta Q^2_{tot}$ and $\Delta E_{tot}$ respectively have been increased for the best possible agreement of the results.

It is evident from the figure that a lower $Q_{min}$ does not substantially influence the shape of the resulting MMFF, but that at least for RHIC kinematics, a lower choice of $Q_{min}$ can be compensated by assuming a different choice of the medium parameters. It is thus not possible to extract definite values for $\hat{q}$ or $D$ from a mode fit to measured single hadron spectra, rather only pairs $(\hat{q}, Q_{min})$ can be determined.  

\section{Comparison with nuclear suppression data}

In this section, we aim at comparing with experimental data. This implies that neither initial position nor initial momentum nor type of the shower initiating parton are known. The probabilities to find a given parton type with given momentum have to be computed in pQCD whereas the probability to produce a parton at a given vertex position can be found from overlap calculations. Note that the need to average over position and initial momentum corresponds to a substantial increase in MC computing time which could not be done without using the scaling laws.

\subsection{The averaging procedure}

We begin the analysis by showing how to compute the nuclear suppression factor $R_{AA}$ using the medium-modified fragmentation function in the hydrodynamically evolving medium. For this, we first have to obtain the single inclusive hard hadron spectrum.

We treat the partonic subprocesses of the hard reaction in leading order pQCD. The straightforward calculation involves the convolution of the initial nucleon \cite{CTEQ1,CTEQ2} (or nuclear \cite{NPDF,EKS98,EPS08}) parton distribution functions with the relevant pQCD subprocesses and yields the single inclusive distribution $\frac{d\sigma^{AB\rightarrow f+X}}{dp_T^2 dy_f}$ of hard partons $f$ in transverse momentum $p_T$ and rapidity $y_f$ where the rest of the reaction $X$ is unobserved (more detailed expressions can be found in \cite{YAS}).

The single inclusive hadron distribution in hadronic momentum $P_T$ and rapidity $y$ follows from the parton spectrum through the convolution with the fragmentation function $D_{f\rightarrow h}(z, \mu_f^2)$ where $z$ is the momentum fraction taken by the hadron and $\mu_f$ is the hadronic momentum scale as

\begin{widetext}
\begin{equation}
\label{E-Fragment}
\frac{d\sigma^{AB\rightarrow h+X}}{dP_T^2 dy} = \sum_f \int dp_T^2 dy_f  
\frac{d\sigma^{AB\rightarrow f+X}}{dp_T^2 dy_f} \int_{z_{min}}^1 dz D_{f\rightarrow h}(z, \mu_f^2)  
\delta\left(m_T^2 - M_T^2(p_T, y_f, z)\right) \delta\left(y - Y(p_T, y_f,z)\right)
\end{equation} 
\end{widetext}
 
with
 
\begin{equation}
M_T^2(p_T, y_f, z) = (zp_T)^2 + M^2 \tanh^2 y_f,
\end{equation}
 
\begin{equation}
z_{min} = \frac{2 m_T}{\sqrt{s}} \cosh y
\end{equation}

and 
\begin{equation}
 Y(p_T, y_f, z) = \text{arsinh} \left(\frac{P_T}{m_T} \sinh y_f \right).
\end{equation}

The nuclear suppression factor is defined as
\begin{equation}
R_{AA}(P_T,y) = \frac{dN^h_{AA}/dP_Tdy}{T_{AA}(0) d\sigma^{pp}/dP_Tdy}
\end{equation}

where $T_{AA}({\bf b})$ is the standard nuclear overlap function. We can compute it by forming the ratio 

\begin{equation}
\label{E-R_AA}
R_{AA}(P_T,y) = \frac{d\tilde\sigma_{medium}^{AA\rightarrow h+X}}{dP_T^2  
dy}/\frac{d\sigma^{pp\rightarrow h+X}}{dP_T^2 dy}
\end{equation}

where ${d\sigma^{pp\rightarrow h+X}}/{dP_T^2 dy}$ follows from Eq.~(\ref{E-Fragment}) when $D_{f\rightarrow h}(z, \mu_f^2)$ is set to be the vacuum fragmentation function whereas ${d\tilde\sigma_{medium}^{AA\rightarrow h+X}}/{dP_T^2 dy}$ is computed from the same equation with $D_{f\rightarrow h}(z, \mu_f^2)$ replace by the suitably averaged MMFF $\langle D_{MM}(z,\mu_f^2)\rangle_{T_{AA}}$. This averaging has to be done over all possible paths of partons through the medium. 

The probability density $P(x_0, y_0)$ for finding a hard vertex at the 
transverse position ${\bf r_0} = (x_0,y_0)$ and impact 
parameter ${\bf b}$ is, again in leading order, given by the product of the nuclear profile functions as
\begin{equation}
\label{E-Profile}
P(x_0,y_0) = \frac{T_{A}({\bf r_0 + b/2}) T_A(\bf r_0 - b/2)}{T_{AA}({\bf b})},
\end{equation}
where the thickness function is given in terms of Woods-Saxon the nuclear density
$\rho_{A}({\bf r},z)$ as $T_{A}({\bf r})=\int dz \rho_{A}({\bf r},z)$. The MMFF must then be averaged over this quantity and all possible directions $\phi$ partons could travel from a vertex as

\begin{widetext}
\begin{equation}
\label{E-P_TAA}
\langle D_{MM}(z,\mu^2) \rangle_{T_{AA}} \negthickspace = \negthickspace \frac{1}{2\pi} \int_0^{2\pi}  
\negthickspace \negthickspace \negthickspace d\phi 
\int_{-\infty}^{\infty} \negthickspace \negthickspace \negthickspace \negthickspace dx_0 
\int_{-\infty}^{\infty} \negthickspace \negthickspace \negthickspace \negthickspace dy_0 P(x_0,y_0)  
D_{MM}(z,\mu^2,\zeta).
\end{equation}
\end{widetext}

Using the approximate scaling relation described in section \ref{S-Scaling}, the medium modified fragmentation function $D_{MM}(z, \mu^2,\zeta)$ for a path $\zeta$ can be found by computing the line integrals $\int d \zeta \hat{q}(\zeta$ or $\int d\zeta D(\zeta)$ over Eqs.~(\ref{E-qhat}),(\ref{E-D}) or by evaluating Eq.~(\ref{E-fmed}) respectively. The MC shower code is then used to compute $D_{MM}(z, \mu^2,\zeta)$ for each value of $\Delta Q^2_{tot}$, $\Delta E_{tot}$ or $f_{med}$ obtained.

As discussed in more detail in \cite{YAS}, there is a conceptual problem with using a MMFF computed for a fixed {\em partonic} scale in Eq.~(\ref{E-Fragment}) where $D(z, \mu_f^2)$ is an object defined at a given {\em hadronic} scale. This is a generic problem in obtaining fragmentations from a MC code which starts with given parton properties, however in practice the scale evolution in the RHIC kinematic range is small as compared to other uncertainties in the computation and the resulting uncertainty can be tolerated. In the following, we use a MMFF determined at the partonic scale $\mu = 20$ GeV.

\subsection{Comparison with data}

With the medium given by the hydrodynamical evolution model described in \cite{Hydro3d} and the expressions for hadron production in vacuum and medium Eq.~(\ref{E-Fragment}), the remaining unknown quantities for a comparison with data are the parameters $K, K_D$ and $K_f$ which link the medium properties in terms of the energy density $\epsilon$ with the parton-medium interaction parameters $\hat{q}, D$ and $f_{med}$. 

Note that according to the results of \ref{S-Qmin} the value of these parameters cannot be uniquely determined for single inclusive hadron spectra, but depends on the choice of the scale $Q_{min}$ in the shower simulation.  In the following, we show the best fit of $K, K_D$ and $K_f$ to the data given the choice $Q_{min} = 1$ GeV.

\begin{figure}[htb]
\epsfig{file=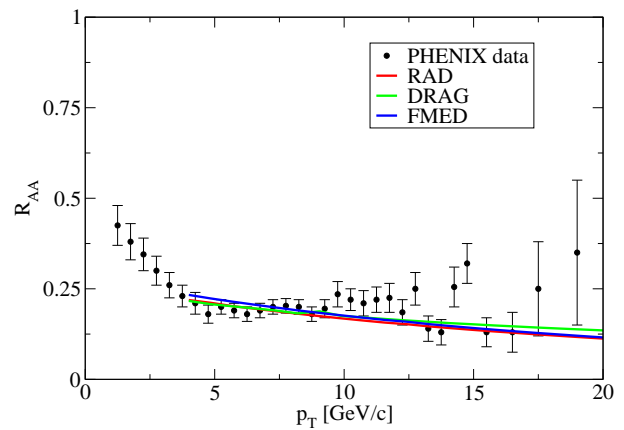, width=8cm}
\caption{\label{F-RAAComp}Calculated nuclear suppression factor for three different scenarios for the parton-medium interaction (see text) as a function of hadron momentum $P_T$ in comparison with data by the PHENIX collaboration \cite{PHENIX_R_AA}. In all cases, the relevant constant relating medium energy density and parton-medium interaction parameter has been fit to data.}
\end{figure}

In Fig.~\ref{F-RAAComp} we show the calculated nuclear suppression factor as a function of hadron momentum $P_T$ for all three scenarios in comparison with the data for $\pi^0$ production in 200 AGeV central Au-Au collisions. The most striking observation is that all three scenarios are surprisingly similar and could not possibly be distinguished by current data for $R_{AA}$. Most notably, all scenarios exhibit a {\em falling} trend as $P_T$ where scenarios based on leading parton energy loss typically exhibit a {\em rising} trend (see e.g. \cite{HydroComp,JetHydro}). In \cite{YAS}, this property was tentatively attributed to the fact that a description of the whole shower keeps track of multiple soft hadron production. It was also suggested that the same physics underlies the enhancement of $dN/d\xi$ in the large $\xi$ region and the falling of $R_{AA}$ with $P_T$. However, the present investigation shows that both ideas must be discarded, as the DRAG scenario in which no enhanced soft hadron production occurs shows also no enhancement of $dN/d\xi$, but the same falling trend of $R_{AA}$ with $P_T$ as the other scenarios. Thus, the falling trend is not a phenomenon characteristic of radiative energy loss but substantially more general. It also has been observed in other models where a modification of the whole shower by the medium was considered, cf. e.g. \cite{HBP,MEFF}.

\begin{figure}[htb]
\epsfig{file=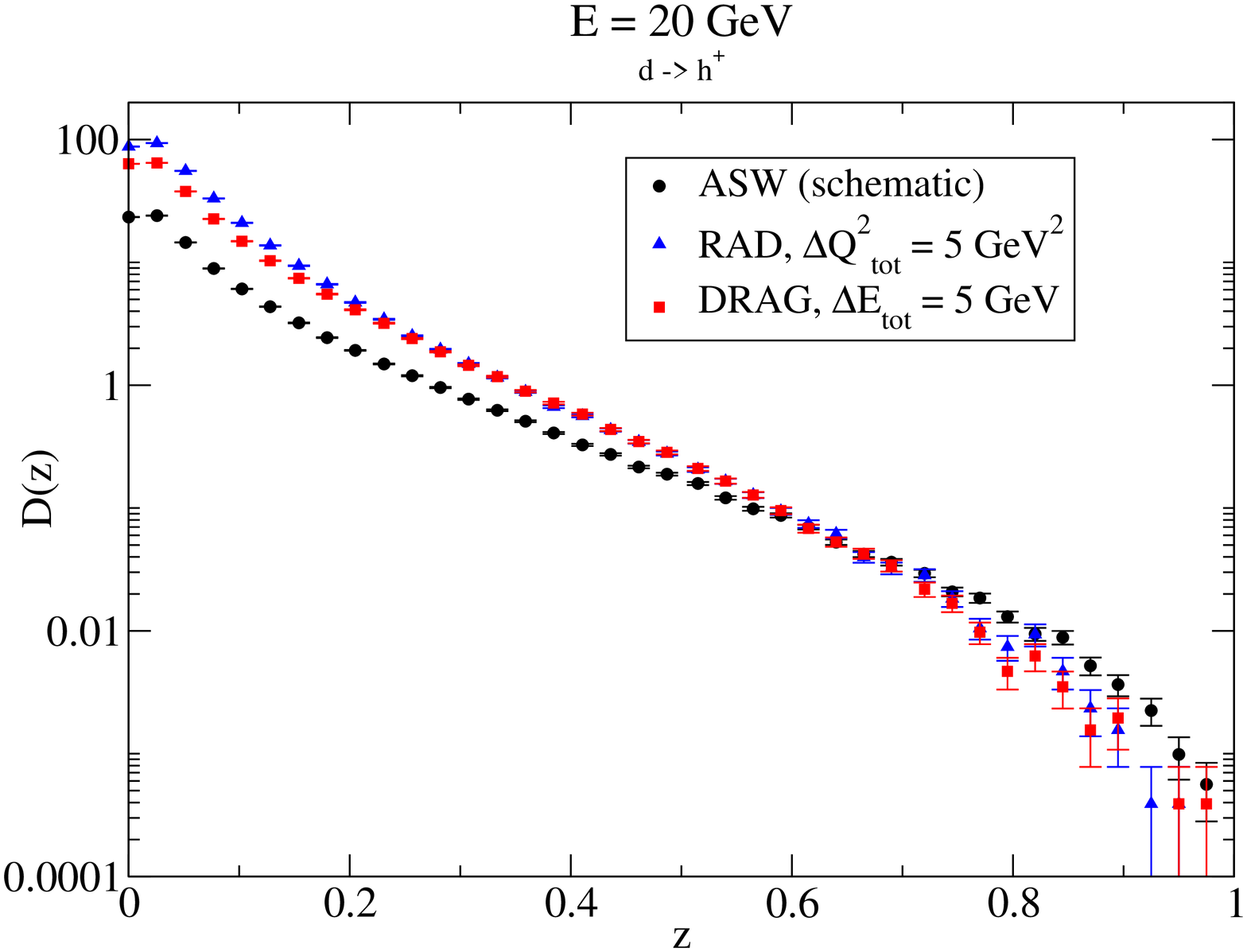, width=8cm}
\caption{\label{F-ASWComp}Comparison of the MMFF of a 20 GeV $d$-quark into charged hadrons for a schematic ASW leading parton energy loss scenario and two different scenarios in which the whole shower is evolves (see text). All parameters are chosen such that the curves agree at $z=0.6$.}
\end{figure}

In order to gain greater insight into the differences between scenarios which compute energy loss for the leading parton and between those where the whole shower  evolution is modified by the medium, we present a schematic comparison between the MMFFs in the RAD, the DRAG and the ASW scenario in Fig.~\ref{F-ASWComp} (the FMED scenario, being in essence indistinguishable from the RAD scenario is not shown here).

It has been pointed out repeatedly (see e.g. \cite{Dihadron,Fragility}) that the ASW scenario applied to RHIC kinematics in essence leads to complete absorption of about 75\% of all partons, $\sim$ 15\% emerge without any energy loss and only a small fraction is found after finite energy loss. To good approximation, the MMFF in the ASW scenario is thus just a downward shift of the vacuum baseline.

This has been done in Fig.~\ref{F-ASWComp} where all parameters have been adjusted such that the curves agree at $z=0.6$. It is obvious that the shape of the schematic ASW result is quite different fron the other scenarios. In particular, the difference between RAD and DRAG is much less pronounced than between either of those and ASW. It is in essence given by the presence or absence of soft hadron production and confined to the region $z<0.2$. Thus, it appears that the different curvature of $D(z)$ at $z>0.5$ is responsible for the rising vs. falling trend in $R_{AA}$, and thus the way the high $P_T$ end of the shower evolves rather than low $P_T$ hadron production are seen in the data.

At present, the falling trend seems not to be supported by the data. Should this be confirmed by more precise measurements, presumably the possibility of complete absorptions of partons by the medium needs to be introduced into the simulation of in-medium shower evolution.

\subsection{Extracting medium parameters}

We can use the above results to tentatively extract medium properties. In the RAD scenario, $K$ is a dimensionless parameter and from the fit shown in Fig.~\ref{F-RAAComp} the value $K=3$ is found. This differs from the result in \cite{YAS} where $K=1.5$ was obtained, note however that in the present work a randomziation of the formation time (see EQ.~(\ref{E-RLifetime})) has been performed and that the effect of the medium prior to thermalization has been included in a schematic way. These two differences account for the changed value of $K$.

With this value of $K$, $\hat{q}_0$, i.e. the highest transport coefficient reached in the evolution in the medium center at thermalization time of 0.6 fm/c is found to be 15.6 GeV$^2$/fm when $Q_{min} = 1 $ GeV is assumed. For $Q_{min} = 0.7$ GeV, the extracted value of$K$ changes to 1.4 and $\hat{q}_0 = 7.2$ GeV$^2$/fm.

$K_D$ is a dimensionful parameter which can be expressed in units GeV$^{-1}$. The same fit yields (due to the numerical coincidence mentioned before) $D_0 = 15.6$ GeV/fm for $Q_{min} = 1$ GeV and $D_0 = 7.2$ GeV/fm for $Q_{min} = 0.7$ GeV. 

While these numbers appear large, it has to be remembered that they reflect a snapshot of the medium at its peak density, from which the energy density drops rapidly as a function of time due to the expansion. Since $f_{med}$ is not in connected to any microscopical properties of the medium, we refrain from analyzing its value here.

\section{Discussion}

We have presented a comparison study of three different mechanisms for the parton-medium interaction in the framework of an in-medium shower evolution. In this study, we have considered a variety of assumptions about the evolution of the medium, among them a constant medium with different length $L$, an evolving medium for parths from the medium center, paths from the medium surface and an average over all possible paths in the medium. We have considered three different types of shower initiators --- heavy quarks, light quarks and gluons. We have studied single parton observables such as the distribution of the leading parton momentum or the energy loss probability density $P(\Delta E)$, single hadron observables like $R_{AA}$ as well as multihadron observables such as the hump-backed plateau $dN/d\xi$. In addition, we have also studied merely technical aspects of modelling such as the role of the cutoff parameter $Q_{min}$ or the effect of randomizing the formation time of partons in branching. From the results in all these different situations, some generic properties can be identified.

$\bullet$ The nuclear suppression factor $R_{AA}$ is not a good observable to distinguish different scenarios of the microscopical interaction of partons with the medium. This statement has been made previously from different angles (see e.g. \cite{HydroComp, Gamma-h, Inversion}) and the present results merely confirm previous findings in yet another framework. More differential observables are needed to determine the nature of parton-medium interaction.

$\bullet$ The falling trend of $R_{AA}$ as a function of $P_T$ is apparently unrelated to low $P_T$ multi hadron production and rather a generic feature observed in models which do not consider energy loss from a leading parton but rather a modification of the whole shower. For example, the results of the Higher Twist approach applied to the leading parton show a rising trend of $R_{AA}$ with $P_T$ \cite{HT1}, however when resummed in the shower evolution equations and applied to the whole shower, the Higher Twist approach leads to a falling trend \cite{HydroComp}, i.e. the same observation is made in quite a different framework. If future data confirm a rising trend, non-trivial modifications to the shower evolution codes, such as the possibility of complete parton absorption by the medium, need to be considered.

$\bullet$ The properties of medium-induced radiation as a mechanism for the parton-medium interaction appear rather generic. There is no observable in this study in which the RAD and the FMED scenarios lead to substantially different results. The useful implication would be that in many observables it is really the underlying physics mechanism one is probing, not technical details of how this mechanism is implemented in a particular model.
 
$\bullet$ In contrast, a different physics mechanism as exemplified here by the DRAG scenario appears distinct in several quantities. Not only is its excitation function in terms of mean energy loss as a function of medium density different than for radiative scenarios (which could be tested by variations in collision centrality), but also the absence of soft hadron production induces pronounced effects in jet observables such as the angular distribution of hadrons around the jet axis or the hump-backed plateau. However, the need to identify a jet in a heavy-ion collision above the soft background introduces additional complications. In essence, a medium-modified jet has properties different from a jet in vacuum and is hence less likely to be identified as jet. A measurement of jets  must be carefully designed to avoid this trigger bias which tends to hide the very effect one would like to study \cite{YAS2}.

$\bullet$ Technical aspects of the modelling, such as the choice of $Q_{min}$ or the randomization of the formation time as investigated here, do not appear to change the results qualitatively. However, there is a substantial ambiguity once one tries to extract quantitative medium parameters from the computation, especially when this extaction is based on a single observable. As is the case for vacuum shower codes, the relevant technical model parameters should eventually be determined by the best fit to a large body of data. 

There are several more properties of the parton-medium interaction which could be exploited to distinguish different scenarios. A very promising candidate is the pathlength dependence of the medium effect. Experimentally, this can be varied moderately by considering the nuclear suppression as a function of the angle of hard hadron with the reaction plane \cite{PHENIX-RP} or more strongly by considering back-to-back correlations (which however require a careful modelling, as the relevant geometry arises as a complicated function of the geometrical bias of the energy loss on the trigger hadron itself \cite{Dihadron1,Dihadron2}. For example, in \cite{Elastic} is was argued based on the different pathlength dependence that elastic energy loss cannot be responsible for the suppression of hard back-to-back hadron correlations.

In the present paper, we have refrained from making any comparison based on pathlength dependence. Such a study (which is quite substantial on its own) along with a comparison with the experimental results on back-to-back correlations and the variation of the suppression as a function of the reaction plane angle will be the topic of a future publication.

\section{Outlook}

The study presented here shows that jet observables are more powerful in order to distinguish different microscopical physics process of the parton-medium interaction than observables which are only sensitive to the leading hadron. However, the need to identify a jet above the background medium may quickly eliminate this advantage, at which point one has to resort to what has been termed the 'golden channel' --- $\gamma$-jet correlations in which the presence of a photon not only allows to get an unbiased jet sample but also reveals the kinematic of the jet. Unfortunately, due to the smallness of the electromagnetic coupling, the statistics in this channel is poor and the measurement is difficult.

Future jet measurements at RHIC may overcome this problem by high luminosity, whereas future measurements at LHC where the scale separation between a hard process and the soft medium is considerably larger than at RHIC may not suffer significantly from trigger bias at all.

There is now good reason to assume that jet observables will reveal important information about the microscopical properties of the medium, and Monte Carlo simulations of in-medium showers such as YaJEM or JEWEL will most likely be the appropriate tools to extract this information.

\begin{acknowledgments}
I'd like to thank Kari J. Eskola for valuable discussions on the problem. This work was financially supported by the Academy of Finland, Project 115262. 
\end{acknowledgments}

\end{document}